\documentclass[%
 reprint,
 superscriptaddress,
preprintnumbers,
nofootinbib,
 amsmath,amssymb,
 aps,
 prc
]{revtex4-1}

\usepackage{dcolumn}

\usepackage{natbib}
\usepackage{mathptmx} 
\usepackage{pstricks}
\usepackage{array}
\usepackage{tabularx}
\usepackage{graphicx}
\usepackage{color,soul}
\usepackage{amsmath,bm}
\usepackage{amssymb}
\usepackage{mathrsfs}
\usepackage{psfrag}
\usepackage{textcomp}
\usepackage{longtable}
\usepackage{multirow}
\usepackage{braket}
\usepackage[mathscr]{euscript}
\usepackage{dcolumn}
\usepackage[inline]{enumitem}

\definecolor{LinkColor}{HTML}{2F348F}
\usepackage{hyperref}
\hypersetup{colorlinks=true, urlcolor=LinkColor, citecolor=LinkColor, linkcolor=LinkColor, anchorcolor=LinkColor, filecolor=LinkColor, menucolor=LinkColor, runcolor=LinkColor, breaklinks=true}

\usepackage{commath}
\usepackage[]{units}
\usepackage{acronym}
\usepackage{bm}
\usepackage{breqn}

\newcommand{\citeref}[1]{Ref.\ \cite{#1}}

\newcommand{\meas}[2]{\unit[${#1}$]{#2}}

\newcommand{\nuc}[2]{$^{#1}${#2}}

\newcommand{\nuccharge}[3]{$^{#1}${#2}$^{(#3)}$}

\newcommand{\rxn}[6]{$^{#1}${#2}$(#3, #4)^{#5}${#6}}
\newcommand{\of}[1]{(#1)}
\newcommand{\ppar}[1] {\left( {#1} \right)}
\newcommand{\brkt}[1] {\left[ {#1} \right]}
\newcommand{\cbrkt}[1]{\left\lbrace {#1} \right\rbrace}
\newcommand{\ee}[1]{\times 10^{#1}}
\newcommand{\wg}{\omega \gamma}

\newcommand{\kpg}{\rxn{38}{K}{p}{\gamma}{39}{Ca}}

\newcommand{\fracline}[2]{\left. {#1} \middle/ {#2} \right.}

\newcommand{\ebeamA}{\meas{15.58}{MeV}}
\newcommand{\ebeamB}{\meas{20.56}{MeV}}
\newcommand{\ebeamC}{\meas{27.17}{MeV}}
\newcommand{\gaus}[3]{
	\frac{
		e^{\ppar{#1 - #2}^2 / \ppar{2 #3}}
	}{
		\sqrt{2\pi} #3
	}
}
\newcommand{\pois}[2]{
	\frac{ 
		\ppar{#2}^#1
	}{
		#1!
	} e^{#2}
}

\newcommand{\eqnref}[1]{Eq.\ \eqref{#1}}
\newcommand{\tableref}[1]{Table~\ref{#1}}

\newcommand{\triumf}  {\textsc {Triumf}}
\newcommand{\geant}[1]{\textsc {Geant{#1}}}

\newcommand{\executeiffilenewer}[3]{%
\if{\Filemodnewest[1]{#1}{#2} == #2}
{\immediate\write18{#3}}\fi%
}

\newacro{VME}  [\MakeUppercase{vme}]{V\MakeUppercase{ersa}module Eurocard}
\newacro{CAMAC}[\MakeUppercase{camac}]{computer automated measurement and control}
\newacro{NIM}  [\MakeUppercase{nim}]{nuclear instrumentation module}
\newacro{ECL}  [\MakeUppercase{ecl}]{emitter coupled logic}
\newacro{FPGA} [\MakeUppercase{fpga}]{field-programmable gate array}
\newacro{DAQ}  [\MakeUppercase{daq}]{data acquisition}
\newacro{DRAGON}[\MakeUppercase{dragon}]{the Detector of Recoils and Gammas of Nuclear Reactions}
\newacro{DSSSD}[\MakeUppercase{dsssd}]{double-sided silicon strip detector}
\newacro{MCP}  [\MakeUppercase{mcp}]{microchannel plate}
\newacro{IC}   [\MakeUppercase{ic}]{ionization chamber}
\newacro{TOF}  [\MakeUppercase{tof}]{time of flight}
\newacro{BGO}  [\MakeUppercase{bgo}]{bismuth germanate}
\newacro{TDC}  [\MakeUppercase{tdc}]{time to digital converter}
\newacro{QDC}  [\MakeUppercase{qdc}]{charge to digital converter}
\newacro{ADC}  [\MakeUppercase{adc}]{amplitude to digital converter}
\newacro{CFD}  [\MakeUppercase{cfd}]{constant fraction discriminator}
\newacro{FIFO} [\MakeUppercase{fifo}]{first in, first out}
\newacro{TSC}  [\MakeUppercase{tsc}]{timestamp counter}
\newacro{IIS}  [\MakeUppercase{iis}]{ion-implanted silicon}
\newacro{NaI}  [\MakeUppercase{n}a\MakeUppercase{i}]{sodium iodide}
\newacro{HPGe} [\MakeUppercase{hpg}{e}]{high purity germanium}
\newacro{MIDAS}[\MakeUppercase{midas}]{Maximum Integrated Data Acquisition System}
\newacro{RFQ}  [\MakeUppercase{rfq}]{radio frequency quadrupole}
\newacro{ECRIS}[\MakeUppercase{ecr}]{electron cyclotron resonance}
\newacro{NLL}  [\MakeUppercase{nll}]{negative log-likelihood}

\graphicspath{{./figures/}}
\renewcommand{\meas}[2]{$#1$ {#2}}

\newcommand{\gammaRayD}{$\gamma$-ray}

\newcommand{\Abr}[2]{
	$\nicefrac{
		X_{\mathrm{#1}}
	}{
		X_{\mathrm{#2}}
	}$
}

\newcommand{\triumfaddr}{TRIUMF, Vancouver, BC V6T~2A3, Canada}
\newcommand{\york}{Department of Physics, University of York, Heslington, York YO10 5DD, UK}
\newcommand{\csm}{Department of Physics, Colorado School of Mines, Golden, CO 80401, USA}
\newcommand{\surrey}{Department of Physics, University of Surrey, Guildford, GU2~7XH, UK}

\newcommand{\mcgillphys}{Department of Physics, McGill University, Montr\'{e}al, QC H3A~2T8, Canada}
\newcommand{\mcmasterphys}{Department of Physics \& Astronomy, McMaster University, Hamilton, ON L8S~4M1, Canada}
\newcommand{\tamuaddr}{Department of Physics \& Astronomy, Cyclotron Institute, and Nuclear Solutions Institute, Texas A\&M University, College Station, TX 77843, USA}

\begin{document}

\author{G. Christian}
\thanks{\href{mailto:gchristian@tamu.edu}{gchristian@tamu.edu}}
\affiliation{\tamuaddr{} }
\affiliation{\triumfaddr{} }

\author{G. Lotay}
\thanks{\href{mailto:g.lotay@surrey.ac.uk}{g.lotay@surrey.ac.uk}}
\affiliation{\surrey{} }

\author{C. Ruiz}
\affiliation{\triumfaddr{} }

\author{C. Akers}
\thanks{Present address: Rare Isotope Science Project, Institute for Basic Science, Daejeon 305-811, Republic of Korea}
\affiliation{\triumfaddr{} }
\affiliation{\york{} }

\author{D.~S. Burke}
\affiliation{\mcmasterphys{} }

\author{W.~N. Catford}
\affiliation{\surrey{} }

\author{A.~A. Chen}
\affiliation{\mcmasterphys{} }

\author{D. Connolly}
\thanks{Present address: TRIUMF, Vancouver, BC V6T 2A3 Canada}
\affiliation{\csm{} }

\author{B. Davids}
\affiliation{\triumfaddr{} }

\author{J. Fallis}
\thanks{Present address: College of Arts and Sciences, North Island College, Courtenay, BC V9N 8N6 Canada}
\affiliation{\triumfaddr{} }

\author{U. Hager}
\affiliation{\csm{} }

\author{D. Hutcheon}
\affiliation{\triumfaddr{} }

\author{A. Mahl}
\affiliation{\csm{} }

\author{A. Rojas}
\affiliation{\triumfaddr{} }

\author{X. Sun}
\thanks{Present address: Division of Physics, Mathematics and Astronomy, California Institute of Technology, Pasedena, CA 91125 USA}
\affiliation{\triumfaddr{} }
\affiliation{\mcgillphys{} }

\title{
	Direct Measurement of Astrophysically Important Resonances in $^{38}\mathrm{K}(p, \gamma){}^{39}\mathrm{Ca}$}
\date{\today}

\begin{abstract}
\begin{description}
\item[Background]
		Classical novae are cataclysmic nuclear explosions occurring when a white dwarf in a binary system accretes hydrogen-rich material from its companion star. Novae are partially responsible for the galactic synthesis of a variety of nuclides up to the calcium ($A \sim 40$) region of the nuclear chart. Although the structure and dynamics of novae are thought to be relatively well understood, the predicted abundances of elements near the nucleosynthesis endpoint, in particular Ar and Ca, appear to sometimes be in disagreement with astronomical observations of the spectra of nova ejecta.
	\item[Purpose]
		One possible source of the discrepancies between model predictions and astronomical observations is nuclear reaction data. Most reaction rates near the nova endpoint are estimated only from statistical model calculations, which carry large uncertainties. For certain key reactions, these rate uncertainties translate into large uncertainties in nucleosynthesis predictions. In particular, the $^{38}\mathrm{K}\left( p, \gamma \right){}^{39}\mathrm{Ca}$ reaction has been identified as having a significant influence on Ar, K, and Ca production. In order to constrain the rate of this reaction, we have performed a direct measurement of the strengths of three candidate $\ell = 0$ resonances within the Gamow window for nova burning, at $386 \pm 10$~keV, $515 \pm 10$~keV, and $689 \pm 10$~keV.
\item[Method]
		The experiment was performed in inverse kinematics using a  beam of unstable $^{38}\mathrm{K}$ impinged on a windowless hydrogen gas target. The $^{39}\mathrm{Ca}$ recoils and prompt $\gamma$ rays from $^{38}\mathrm{K}\left( p, \gamma \right){}^{39}\mathrm{Ca}$ reactions were detected in coincidence using a recoil mass separator and a bismuth-germanate scintillator array, respectively.
\item[Results]
		For the $689$ keV resonance, we observed a clear recoil-$\gamma$ coincidence signal and extracted resonance strength and energy values of $120^{+50}_{-30}~\mathrm{(stat.)}^{+20}_{-60}~\mathrm{(sys.)}~\mathrm{meV}$ and $679^{+2}_{-1}~\mathrm{(stat.)} \pm 1~\mathrm{(sys.)}~\mathrm{keV}$, respectively. We also performed a singles analysis of the recoil data alone, extracting a resonance strength of $120 \pm 20~\mathrm{(stat.)} \pm 15~\mathrm{(sys.)}$~meV, consistent with the coincidence result. For the $386$ keV and $515$ keV resonances, we extract $90\%$ confidence level upper limits of $2.54$ meV and $18.4$ meV, respectively.
\item[Conclusions]
		We have established a new recommended $^{38}\mathrm{K}(p, \gamma){}^{39}\mathrm{Ca}$ rate based on experimental information, which reduces overall uncertainties near the peak temperatures of nova burning by a factor of $\sim 250$. Using the rate obtained in this work in model calculations of the hottest oxygen-neon novae reduces overall uncertainties on Ar, K, and Ca synthesis to factors of $15$ or less in all cases.
\end{description}
\end{abstract}

\maketitle

\section{Introduction}

Classical novae are some of the most common explosive stellar events to occur in our galaxy, with an estimated frequency of  $35\pm 11$ per year \cite{Shafter1997}.  Novae happen when a white dwarf in a binary system  accretes hydrogen-rich material from its main-sequence companion, igniting thermonuclear runaway. Observations of the spectra of ejected material indicate that two main classes of nova exist, depending on the initial composition of the underlying white dwarf: carbon-oxygen (CO) and oxygen-neon (ONe).  Model calculations indicate that ONe novae, which occur on more massive white dwarves, can reach peak temperatures around \meas{0.4}{GK} and synthesize nuclei up to the calcium region ($A \sim 40$). At present, there are a number of outstanding discrepancies between astronomical observations of the spectra of nova ejecta \cite{Pottasch1959,Andrea1994,Arkhipova2000,Evans2003} and nova model predictions \cite{Jose2006, Starrfield2009}. In particular, the model predictions of \citeref{Starrfield2009} indicate Ar and Ca abundances at roughly the solar level, while in contrast the observations of \citeref{Andrea1994} point towards nova ejecta with Ar and Ca abundances around an order of magnitude greater than solar. Resolution of such discrepancies requires that nova models be capable of making detailed predictions regarding the synthesis of nuclides in the Ar--Ca region. In turn, this requires improved constraints on the rates of key nuclear reactions involved in nova nucleosynthesisis, in particular for reactions near the nucleosynthesis endpoint.

In 2002, Iliadis \emph{et al}.\ published a seminal paper investigating the influence of nuclear reaction rate variations on nucleosynthesis in classical novae \cite{Iliadis2002}. In this study, the authors varied the rates of $64$ nuclear reactions within their recommended uncertainties and examined the effect of these variations on the  nucleosynthesis predictions of seven different nova models. For the hottest model included in the study, reaching a peak temperature of \meas{0.418}{GK}, the authors identified the \kpg{} reaction as having a significant influence on the production of Ar, K, and Ca. Qualitatively, the predicted abundances of these elements were found to vary by respective factors of $24$, $58$, and $57$ when the \kpg{} rate was varied within its existing uncertainties. When \citeref{Iliadis2002} was published, the \kpg{} rate was estimated entirely from statistical model predictions with no experimental nuclear physics input \cite{Iliadis2001}. This rate estimate was assigned an overall uncertainty of $10^{4}$, i.e.\ the upper and lower limits were established at $100$ and $0.01$ times the central value, respectively. The importance of this reaction for nova nucleosynthesis, along with the paucity of experimental input regarding the accepted rate, prompted an attempt by the present authors to measure the strengths of the three $\ell = 0$ resonances lying within the Gamow window for ONe novae ($T_{\mathrm{peak}} \simeq 0.2$--\meas{0.4}{GK}). The first results of this study were published in a review article \cite{BDreview} and a recent Letter, which recommends a new, experimentally-based rate with uncertainties over two orders of magnitude smaller than before \cite{PhysRevLett.116.132701}. In the present Article, we expand upon \citeref{PhysRevLett.116.132701}, providing significantly more detail concerning the experiment and data analysis. We also report the results of a new sensitivity study investigating the effect of our measurement on the synthesis of Ar, K, and Ca in classical novae. The results presented here supersede those published previously.

\section{Experiment}
\label{sec:Exp}

The experiment was performed in the ISAC-I \cite{Laxdal2003400} hall at \triumf{}, Canada's national laboratory for particle and nuclear physics. A beam of radioactive \nuc{38}{K} was produced by impinging \meas{500}{MeV} protons from the \triumf{} cyclotron onto a high-power TiC production target. The \nuccharge{38}{K}{1+} ions produced by spallation reactions in the target were extracted and sent through a high-resolution mass separator. They were then charge bred to the $7^+$ charge state in an \ac{ECRIS} charge state booster before post-acceleration. The charge breeding is necessary because the ISAC-I \ac{RFQ} is restricted to a mass-to-charge ratio of $30$ or less \cite{csb}. 
	
The \nuccharge{38}{K}{7+} beam was delivered to \ac{DRAGON} where it impinged on a windowless extended gas target \cite{Hutcheon2003190}, filled with H$_2$ at an average pressure and temperature of \meas{10.6}{mbar} and \meas{298}{Kelvin}, respectively. The H$_2$ was cleaned by continuous recirculation through a LN$_2$ cooled zeolite trap. The prompt $\gamma$ rays from \kpg{} reactions were detected in array of $30$ \ac{BGO} scintillators surrounding the target, while the \nuc{39}{Ca} recoils were transmitted to the focal plane of DRAGON, separating them from unreacted and elastically scattered \nuc{38}{K}. A timing signature for recoils was established as the time difference between signals from a pair of \acp{MCP} separated by \meas{59}{cm}, which detected secondary electrons produced by the interaction of the recoil ions with a diamond-like carbon foil intersecting the beam line. The total kinetic energy and stopping power of the recoil ions was measured in a multi-anode \ac{IC} \cite{Vockenhuber2009372}. Coincidences between recoils and prompt $\gamma$ rays were identified using a timestamp-based algorithm \cite{dragondaq}. The \nuc{39}{Ca} recoils were separated from a background of scattered and charge-changed \nuc{38}{K} (``leaky beam'') based primarily on the local \ac{TOF} between the two \acp{MCP} (``\ac{MCP} \ac{TOF}'') and the time difference between the $\gamma$ ray and the upstream \ac{MCP} (``separator \ac{TOF}'').

Laboratory beam energies of \ebeamA{}, \ebeamB{}, and \ebeamC{} were employed for measurements of the $386 \pm 10$~keV, $515 \pm 10$~keV, and $689 \pm 10$~keV resonances, respectively. The beam energies were measured using the procedure given in \citeref{Hutcheon201270}. The beam was centered on \meas{2}{mm} slits downstream of \ac{DRAGON}'s first magnetic dipole, and the measured field value was converted to energy by solving the relativistically-correct equation,
\begin{equation}
\label{eq:Field2Energy}
	E/A = c_{\mathrm{mag}} \left( qB/A \right)^2 - \frac{1}{2uc^2}\left(E/A \right)^2,
\end{equation}
	where $E$, $A$, and $q$ are the beam kinetic energy, mass number, and charge state, respectively, and $u$ is the atomic mass unit. The quantity $c_{\mathrm{mag}}$ is related to the effective bending radius of the dipole. The recommended value from \citeref{Hutcheon201270}, \meas{c_{\mathrm{mag}} = 48.15 \pm 0.07}{MeV$\cdot$T$^2$}, was employed for this experiment. The estimated uncertainty on this procedure is $0.17\%.$

The chosen beam energies cover respective center-of-mass energies in the \ac{DRAGON} gas target of $386 \pm 13$, $515 \pm 13$, and \meas{689 \pm 13}{keV}. The resonances in question were previously identified as $5/2^{+}$ \nuc{39}{Ca} states through \rxn{40}{Ca}{{}^{3}\mathrm{He}}{\alpha}{39}{Ca} \cite{Hinds1966}, \rxn{40}{Ca}{d}{t}{39}{Ca} \cite{Doll1976}, and \rxn{40}{Ca}{p}{d}{39}{Ca}  \cite{Matoba1993} transfer reaction studies. Their recommended excitation energies are $6157 \pm 10$, $6286 \pm 10$, and \meas{6460 \pm 10}{keV}, corresponding to \nuc{38}{K}$ + p$ resonances at  $386 \pm 10$, $515 \pm 10$, and \meas{689 \pm 10}{keV}, respectively \cite{Singh2006}.  The respective $(p,\gamma)$ cone angles for measurements at the \ebeamA{}, \ebeamB{}, and \ebeamC{} beam energies were \meas{5.98}{mrad}, \meas{5.29}{mrad}, and \meas{4.73}{mrad}. Each of these is well within the \meas{\pm 21}{mrad} angular acceptance of \ac{DRAGON} \cite{Ruiz2014}. 

 For each beam energy, only a single charge state was transmitted to the end of \ac{DRAGON}. The respective charge states were $7^{+}$, $9^{+}$, and $10^{+}$ for the \ebeamA{}, \ebeamB{}, and \ebeamC{} beam energies. The charge state fractions and stopping powers for K and Ca ions passing through the gas target were measured separately using stable beams of \nuc{39}{K} and \nuc{44}{Ca}. Charge state fractions were determined by measuring the ratio
$ 
({I_2^{{g}}} / {I_0^{{g}}}) \cdot ({I_0^{{ng}}} / {I_1^{{ng}}}),
$ 
where $I_0$, $I_1$, and $I_2$ represent the current on Faraday cups upstream of the gas target, downstream of the gas target, and downstream of the first magnetic dipole, respectively; and the superscripts $g$ and $ng$ represent currents measured with and without gas in the target, respectively. Current measurements were taken with the magnetic dipole set to accept each of the charge states that resulted in a measurable $I_2$. The resulting distributions were then fit with a Gaussian function (normalized to unity). The value of the Gaussian at each charge state was taken to represent the corresponding charge state fraction. Measurements were taken at three different beam energies spanning the range of beam energies employed in the experiment, and the resulting charge fractions were fit with a quadratic function. The value of this quadratic function at the various beam energies employed in the experiment was then taken as the charge state fraction to use in the recoil yield analysis. All fits were performed using MINUIT and errors on the fit parameters were calculated using MINOS \cite{JAMES1975343}. The errors on the Gaussian fit were propagated along with the errors on the quadratic interpolation to arrive at the final error on the charge state fractions used in the analysis. 

\begin{figure}
\centering
\includegraphics{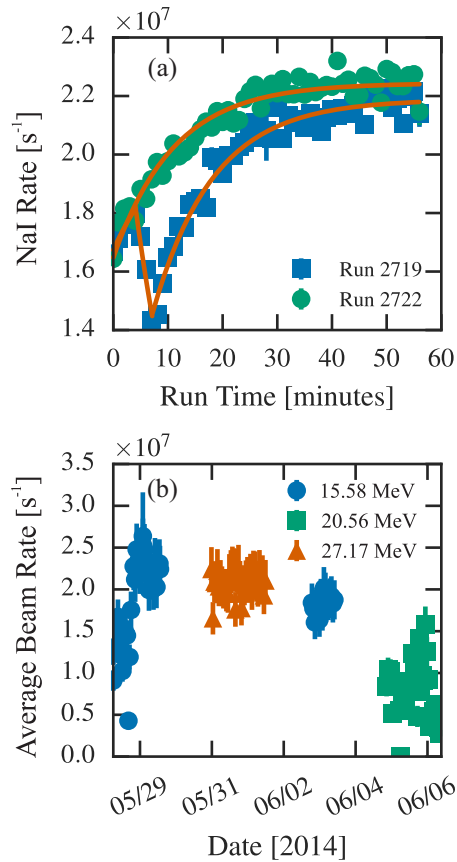}
\caption{
	(a) Sample rate vs.\ time curves measured in the NaI detector. The green filled circles correspond to a run where the beam rate was constant, while the blue filled squares correspond to a run where there were significant changes in the rate. The solid orange lines show the fit results used to extract the average beam intensities. (b) Average beam rate determined for each \meas{{\sim}1}{hour} run taken during the experiment. The filled circles, rectangles, and triangles denote each run's beam energy, as indicated in the legend.
}
\label{fig:beamrates}
\end{figure}

The number of incoming \nuccharge{38}{K}{7+} ions was determined by counting delayed ($t_{1/2} = $ \meas{7.6}{minutes}) \meas{2.2}{MeV} $\gamma$ rays emitted by the daughters of beam ions implanted into the mass slits just downstream of DRAGON's first electric dipole. These $\gamma$ rays were detected in a {NaI} scintillator with an efficiency of $(8.46 \pm 0.95) \ee{-6}$. This efficiency was determined from a \geant{4} \cite{Agostinelli2003250} simulation, which included the entire geometry of the mass slit box and {NaI} detector. The $11\%$ relative uncertainty on the NaI efficiency was determined by comparing simulation results to known \nuc{22}{Na} and \nuc{137}{Cs} source measurements. This analysis includes an uncertainty on the source position of \meas{\pm 0.5}{cm}. The average beam rate for each \meas{{\sim}1}{hour} run was determined by fitting the decay rate vs.\ time curves with the expected response function,
\begin{equation}
 	\label{eq:Rate}
 		A(t) = I \left( 1 - e^{-\lambda t} \right) + N_0 \lambda e^{-\lambda t},
\end{equation}
where $A(t)$ is the decay rate, $I$ is the average beam intensity, $N_0$ is the initial number of particles implanted in the slit, and \meas{\lambda = 1.5 \ee{-3}}{s$^{-1}$} is the \nuc{38}{K} decay constant. In the fit, both $I$ and $N_0$ were allowed to vary as free parameters. Cases where the average beam rate fluctuated significantly over the course of a run were identified by a noticeable deviation from the expected response. These fluctuations in the beam rate (or the complete loss of beam delivery) arose from a number of sources upstream of the \ac{DRAGON} target, for example loss of the \meas{500}{MeV} proton beam or Faraday cup readings taken by the ISAC-I operators. In these cases, differing sections of the run were visually identified and independently fit to \eqnref{eq:Rate}. \figref{fig:beamrates}(a) shows sample fitted rate vs.\ time curves for two runs, one with a constant beam rate and the other with a varying beam rate (and corresponding piecewise fit). \figref{fig:beamrates}(b) shows the average beam rates determined for each run throughout the course of the experiment. The overall \nuc{38}{K} rate was approximately \meas{2 \ee {7}}{particles per second}.

\subsection{386(10) keV and 515(10) keV Resonances}

\begin{figure}
\centering
\includegraphics{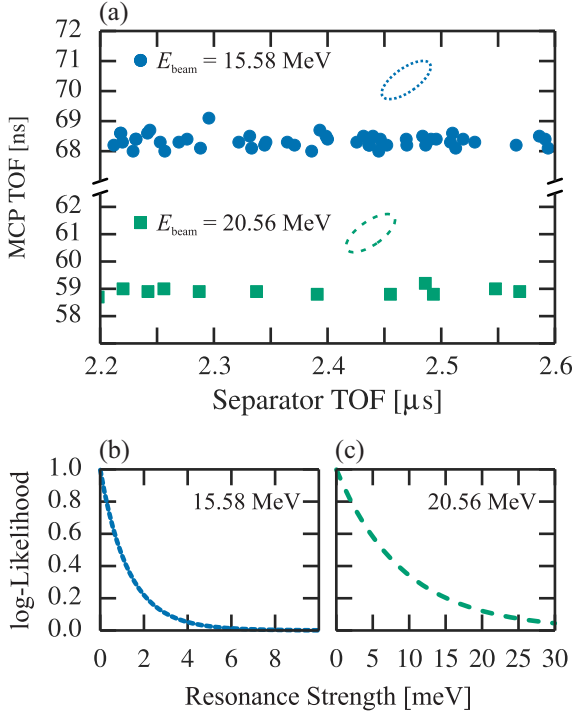}
\caption{
	(a) \ac{MCP} vs.\ separator \ac{TOF} for the \ebeamA{} and \ebeamB{} beam energies. The blue dotted and green dashed ovals represent the expected location of recoils for the \ebeamA{} and \ebeamB{} beam energies, respectively. (b) Profile likelihood curve for the \ebeamA{} beam energy. (c) Profile likelihood curve for the \ebeamB{} beam energy.
}
\label{fig:UL}
\end{figure}

At beam energies of \ebeamA{} and \ebeamB{} (corresponding to the \meas{386 \pm 10}{keV} and \meas{515 \pm 10}{keV} resonances, respectively), we observed zero events in the expected recoil region. This is demonstrated in \figref{fig:UL}(a), which shows \ac{MCP} vs.\ separator \ac{TOF} spectra for each of the \ebeamA{} and \ebeamB{} beam energies. The dashed and dotted ellipses included on the plots indicate the expected location of \nuc{39}{Ca} recoils, based on \geant{3} simulations of the reaction and transmission through the \ac{DRAGON} separator. As can be seen, in both cases no recoil events fall within this expected window.  As a result, we extracted upper limits on the resonance strengths using a modification of the Rolke profile likelihood method for calculating confidence intervals in the presence of uncertain background rates and detection efficiencies \cite{Rolke2005493}.
In the standard Rolke treatment, the likelihood is the product of the individual likelihoods describing the signal rate $\mu$, background rate $b$ (both treated as Poisson), and the detection efficiency $\eta$ (treated as Gaussian with uncertainty $\sigma_\eta$). Mathematically, this is expressed as
	\begin{multline} \label{eq:fullLikelihood}
		\mathscr{L} (\mu, b, \eta \vert x, y, z) = \\
		\brkt{\pois{x}{\eta\mu+b}}
		\brkt{\pois{y}{b\tau}}
		\brkt{\gaus{z}{\eta}{\sigma_\eta}},
	\end{multline}
where $x$ is the number of events observed in the signal region, $y$ is the number of events observed in a background region that is $\tau$ times as large as the signal region, and $z$ is the observed signal rate. Equation~\eqref{eq:fullLikelihood} is then maximized with respect to $b$ and $\eta$ to construct a one-dimensional likelihood curve that is a function of only the signal strength $\mu$ and can be analyzed to extract upper limits.

In the present analysis, we extend the Rolke method to also account for uncertainties in the resonance energy $E_r$, the number of incoming beam particles $N$, and the \nuc{38}{K} + H$_2$ stopping power $\epsilon$. Each of these quantities factors into the calculation of the resonance strength, and hence their uncertainties should be included for a complete treatment of the problem. For each of these quantities, we treat the uncertainty as Gaussian (with widths $\sigma_{E}$, $\sigma_N$, and $\sigma_\epsilon$, respectively). The complete likelihood function is then given by
\begin{multline} \label{eq:lWg}
	\mathscr {L} \ppar{\wg, b, \eta, E_r, N, \epsilon \vert x, y, z, E_{r0}, N_0, \epsilon_0} =\\ 
	\brkt{\pois{x}{\eta\mu + b}}
	\brkt{\pois{y}{b\tau}}
	\brkt{\gaus{z}{\eta}{\sigma_\eta}}  \times \\
	\brkt{\gaus{E_{r_0}}{E_r}{\sigma_{E}}}
	\brkt{\gaus{N_0}{N}{\sigma_N}}
	\brkt{\gaus{\epsilon_0}{\epsilon}{\sigma_\epsilon}},
\end{multline}
where $E_{r_0}$, $N_0$, and $\epsilon_0$ are the observed central values of the resonance energy, beam ions on target, and stopping power, respectively. In \eqnref{eq:lWg}, the signal rate $\mu$ is no longer a constant parameter but rather a function of the resonance strength $\wg$, resonance energy $E_r$, number of incoming beam particles $N$,  center-of-mass stopping power $\epsilon$, beam mass $M$, and target mass $m$,
\begin{equation} \label{eq:muFunc}
			\mu\of{\wg, E_r, N, \epsilon} = \frac{ 
		N \ppar{\wg}\ppar{hc}^2
	}{
		2 \epsilon \brkt{E_r^2 + 2 E_r m M / \ppar{m+M}}
	}.
\end{equation}
Following the Rolke prescription, we maximize \eqnref{eq:lWg} with respect to the ``nuisance'' parameters $\cbrkt{b, \eta, E_r, N, \epsilon}$ to arrive at a profile likelihood that is a function of only the resonance strength $\wg$. In practice, we first take the negative logarithm of \eqnref{eq:lWg} and then calculate the minimum numerically using the MINUIT package \cite{JAMES1975343}. The resulting profile likelihoods for the \ebeamA{} and \ebeamB{} beam energies are shown in Figures~\ref{fig:UL}(b)~and~\ref{fig:UL}(c), respectively (plotted as negative log-likelihoods). To extract single-sided $68\%$, $90\%$, and $95\%$ upper limits from the profile likelihood curves, we follow exactly the prescriptions of Ref.~\cite{Rolke2005493}. The resulting upper limits, along with all of the measured parameters going into the upper limit calculation are summarized in Tables~\ref{table:upperlimitsA}~and~\ref{table:upperlimitsB}. It should be noted that when we refer to ``$68\%$'' or ``$95\%$'' confidence intervals, we mean the area under a normalized gaussian distribution between the $\pm 1\sigma$ or $\pm 2\sigma$ limits. These are more precisely equal to $68.27\%$ and $95.45\%$, respectively.


\begin{table}
	\caption{
		Summary of observed quantities going into the resonance strength upper limit calculations, for the \ebeamA{} beam energy.
	}
	\centering
	\begin{ruledtabular}
		\begin{tabular}{cc}
			{Quantity}				& Value \\
			\hline \\ [-8pt]
			Background rate		& $4.44 \ee{-2}$				\\
			Beam ions on target	& $(2.88 \pm 0.36)\ee{12}$		\\
			Stopping power [eV~cm$^2$]	& $(3.78 \pm 0.14)\ee{-15}$ \\
			Detection efficiency  		& $0.093 \pm 0.016$			\\
			\hline	\\ [-8pt]
			$68\%$ upper limit [meV] 		& $1.16$		\\
            $90\%$ upper limit [meV]		& $2.54$		\\
			$95\%$ upper limit [meV] 		& $3.53$		\\
		\end{tabular}
	\end{ruledtabular}
	\label{table:upperlimitsA}
\end{table}

\begin{table}
	\caption{
		Summary of observed quantities going into the resonance strength upper limit calculations, for the \ebeamB{} beam energy.
	}
	\centering
	\begin{ruledtabular}
		\begin{tabular}{cc}
			{Quantity}				& Value \\
			\hline \\ [-8pt]
            Background rate 	& $4.44 \ee{-3}$ \\
            Beam ions on target	& $(8.8 \pm 1.2)\ee{11}$  	\\
            Stopping power [eV~cm$^2$]	&  $(4.04 \pm 0.14)\ee{-15}$  \\
            Detection efficiency  		&  $0.062 \pm 0.011$ 	\\
			\hline	\\ [-8pt]
            $68\%$ upper limit [meV] 		& $8.59$				\\
            $90\%$ upper limit [meV]		& $18.4$				\\
			$95\%$ upper limit [meV] 		& $25.5$ 				\\
		\end{tabular}
	\end{ruledtabular}
	\label{table:upperlimitsB}
\end{table}

\subsection{689(10) keV Resonance}
\label{ss:Coinc}

\begin{figure*}
\includegraphics
{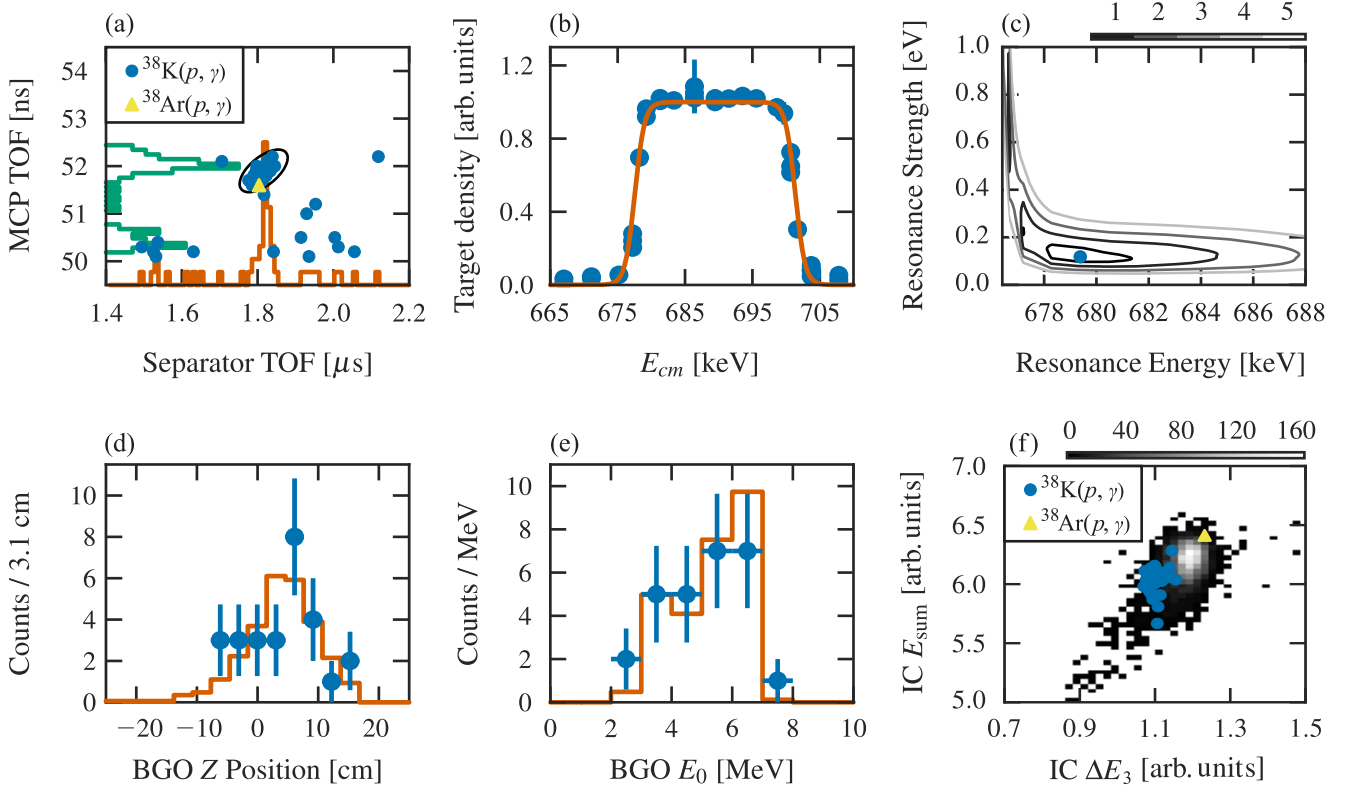}
\caption[]{
\newcommand{\ppp}[1]{{{#1})}}
Summary of the coincidence analysis for the data taken with a beam energy of \ebeamC{}. The individual descriptions of panels (a) through (f) are as follows:
	\ppp{a} Separator vs.\ MCP TOF particle identification spectrum. The blue filled circles represent data collected with the radioactive \nuc{38}{K} beam, while the single filled yellow triangle represents data collected with a \nuc{38}{Ar} beam, for background characterization. The open ellipse outlines the expected recoil region. Projections onto the horizontal and vertical axes are also included (as the unshaded orange and green histograms).
	\ppp{b} Target density as a function of center-of-mass beam energy. The filled circles with error bars represent the data points and the solid line shows the fit to \eqnref{eq:TargetFit}.
	\ppp{c} NLL contour plot, calculated by comparing simulated and measured \ac{BGO} $z$ positions as explained in the text. The solid blue point shows the location of the global minimum.
	\ppp{d} Measured BGO $z$-position distribution for recoil events (filled circles), compared with the best-fit simulation result at $E_r = 679$ keV and $\wg = 120$ eV (solid orange lines).
	\ppp{e} Same as panel (d), but showing the measured energy of the most energetic $\gamma$-ray hit in the BGO array.
	\ppp{f} Total energy deposited in the \ac{IC} vs.\ energy loss in the third (most downstream) anode. The filled (blue) circles show the location of the \nuc{39}{Ca} coincidence recoils observed with the \nuc{38}{K} beam. The filled yellow triangle denotes the location of the event observed with a beam of pure \nuc{38}{Ar}. The greyscale color map shows the location of all heavy-ion singles events observed with the \ebeamC{} \nuc{38}{K} beam. This distribution is dominated by leaky beam.
}
\label{fig:e689coinc}
\end{figure*}


In contrast to the two lower-energy resonances, we observed a clear recoil signal when running with a beam energy of \ebeamC{}. This is demonstrated in the separator TOF vs.\ MCP TOF distribution shown by the filled circles in \figref{fig:e689coinc}(a). This spectrum exhibits a clear clustering of $27$ recoil events in the region indicated by the open ellipse. The  \ac{BGO} $z$-position distribution of the identified recoil events is clustered downstream of the target center, indicating a resonance energy less than the central value of \meas{689}{keV} \cite{Hutcheon201270}. Hence to extract a resonance strength, $\wg,$ and a resonance energy, $E_r$, we use a technique similar to that employed in \citeref{PhysRevC.81.045808}. For a fixed beam energy of \ebeamC{}, we generate a simulated \ac{BGO} $z$-position spectrum over the range of resonance energies contained within the gas target. For the simulations, we use the standard \ac{DRAGON} \geant{3} package \cite{darioThesis} and convolute the resulting \ac{BGO} energies with a realistic hardware threshold. The hardware threshold was determined experimentally by taking long background runs with the threshold set to the value employed in the experiment, and to a reduced value of \meas{50}{mV}. The resulting spectra were normalized, divided into each other, and fit with a Fermi function to arrive at the functional form used in the analysis. Following the threshold convolution, we scale the simulated spectra by the factor
\begin{equation}
\label{eq:SimScale}
	 \eta Y_{\wg} \fracline{N_{b}}{N_{\mathrm{sim}}},
\end{equation}
where $\eta = 0.121 \pm 0.003$ is the heavy-ion detection efficiency,
$Y_{\wg}$ is the reaction yield at a given resonance strength $\wg$,
$N_b = (2.53 \pm 0.30) \ee {12}$ is the number of incoming beam ions, and
$N_{\mathrm{sim}} = 50,000$ is the number of simulated events.
Scaled in this manner, the simulated spectrum represents both the magnitude and the shape of the \ac{BGO} $z$-position distribution for a given $\wg$ and $E_{r}$.

The $\gamma$-ray efficiency is implicitly included in the generation of the simulated spectra since the number of counts appearing in the spectra prior to scaling is determined by the detection efficiency, as modeled in the \textsc{Geant3} simulation. This modeling is sensitive to the branching ratios for $\gamma$-ray decay from the \meas{6460}{keV} state in \nuc{39}{Ca}. These branching ratios have not been measured, and hence we have assumed dominant decays either directly to the ground state or through the first excited $\nicefrac{5}{2}^-$ state, as observed for the decay of known $\nicefrac{5}{2}^+$ excited states in the well-studied mirror nucleus \nuc{39}{K} \cite{Singh2006}. The location of the $\nicefrac{5}{2}_1^-$ state in \nuc{39}{Ca} has not been conclusively assigned, but there are a number of candidates in the ${\sim} 3$--$4$~MeV excitation energy region \cite{Singh2006}. Hence for the present analysis, we have assumed decay through a state at \meas{3.5}{MeV} to represent the feeding through the $\nicefrac{5}{2}_1^-$.  To quantatively account for the uncertainty related to the $\gamma$-ray decay scheme, we have utilized a profile likelihood technique to marginalize over the unknown branching ratios. Specifically, we performed separate simulations for a range of different fractional feedings directly to the ground state or through a state at \meas{3.5}{MeV}. In the simulations, the ground state/excited state ratios ranged from $0\%$--$100\%$ in steps of $10\%$. For each set of simulations, we took the branching with the highest likelihood value and incorporated it into the eventual likelihood surface used to extract confidence intervals on the resonance strength and energy (the calculation of likelihoods and construction of the likelihood surface is detailed later in this section). This technique of using profile likelihoods to marginalize over relevant, but uninteresting ``nuisance'' parameters is well established in the statistical literature; see, for example Refs.~\cite{Rolke2005493, JamesStats}.
 It should be noted that the uncertainty on the $\gamma$-ray detection efficiency is dominated by geometrical and Monte-Carlo uncertainties and not the unknown branching ratios.

The yield parameter in Eq.~\eqref{eq:SimScale}, $Y_{\wg}$, is given by the convolution of the standard Breit-Wigner narrow-resonance cross section \cite{IliadisBook} with the gas target density profile. The density profile was measured in a previous experiment by recording the \gammaRayD{} yield from the $^{3}\mathrm{He}({}^{12}\mathrm{C}, p){}^{14}\mathrm{N}\gamma$ reaction in a shielded \ac{BGO} detector moved along the length of the target \cite{marianoThesis}. These data (scaled to the \ebeamC{} beam energy employed in the present experiment) are shown in \figref{fig:e689coinc}(b). The density profile was determined by fitting the data with the following function:
\begin{equation}
\label{eq:TargetFit}
		f(E) = \fracline{1}{
\left[ {1 + e^{
				\fracline{
					\left(  \left\vert  E - E_0 \right\vert - \fracline{\Delta E}{2} \right)
				}{
					a
				}
		} } \right]} 
,
\end{equation}
where $E_0$ is the beam energy at the center of the gas target, $\Delta E$ is the energy loss across the full length of the gas target, and $a$ is a free parameter. The resulting best-fit is shown as the orange solid line in \figref{fig:e689coinc}(b). The fitting procedure implicitly includes the stopping power, \meas{\epsilon = (3.95 \pm 0.14)\ee{-15}}{eV~cm$^{2}$} (in the center-of-mass frame).


To extract a resonance strength and energy, we calculate the \ac{NLL} by comparing  our model (the scaled \ac{BGO} $z$-position simulations) with experimental data, over a grid of resonance strengths and energies. We assume the counts per bin in the \ac{BGO} $z$-position spectra are Poisson distributed, meaning the \ac{NLL} is given by
\begin{equation}
\label{eq:Likelihood}
	- \ln \mathscr{L} =
\sum_{i} \left \lbrace \ln  \left( {n_i}! \right) - n_i \ln \left( f_i \right) \right \rbrace   + S.
\end{equation}
Here $n_i$ is the number of measured counts in bin $i$, $f_i$ is the number of simulation counts in bin $i$, and $S$ is the integral of the simulated distribution. The result of this likelihood analysis is shown in \figref{fig:e689coinc}(c). This figure shows a contour plot of the \ac{NLL} as a function of the resonance energy and resonance strength, which contains two local minima. The first (global) minimum is in the constant-pressure region of the target with \meas{E_r = 679}{keV}, \meas{\wg = 120}{meV}, and $- \ln \mathscr{L}_0 = 16.2$. The second (local) minimum is far upstream in the target, where the density has not yet reached equilibrium, at \meas{E_r = 677}{keV}, \meas{\wg = 650}{meV}, and $- \ln \mathscr{L}_1 = 16.9$. Based on the \ac{NLL} values, we exclude the  $E_r = 677~${}keV solution at a $76\%$ significance level. This significance level was calculated using the likelihood ratio test, wherein $2 \ln [ {\mathscr{L}_0}/{\mathscr{L}_1} ]$ (here equal to $1.4$) is taken to be $\chi_1^2$ distributed \cite{JamesStats}.  The significance level is thus the value of $X_1^2 (1.4)$, where $X_1^2(x)$ is the $\chi^2$ cumulative distribution function with one degree of freedom.
The resulting best fits to both the \ac{BGO} $z$-position and the $\gamma$-ray energy spectra are shown in Figures~\ref{fig:e689coinc}(d)~and~\ref{fig:e689coinc}(e), respectively.

\begin{table}
	\caption{
			Sources of systematic uncertainty for the measurements at \ebeamC{} beam energy.
	}
	\centering
	\begin{ruledtabular}
		\begin{tabular}{ccc}
			\multirow{2}{*}{Quantity}	& 
            \multirow{2}{*}{Measured Value}	&
            {Relative} \\
			& & {Uncertainty} \\
			\hline \\ [-8pt]
			\nuc{38}{Ar} background & (see text) & $^{+0\%}_{-50\%}$  \\
			Beam ions on target  & $(2.53 \pm 0.30) \ee{12}$ & $12\%$              \\
			BGO efficiency & $0.541 \pm 0.054$ & $10\%$                     \\
			Stopping power [eV~cm$^2$] & $(3.95 \pm 0.14)\ee{-15}$ & $3.5\%$  \\
			\ac{MCP} transmission & $0.789 \pm 0.021$ & $2.7\%$               \\
			Charge state fraction & $0.192 \pm 0.002$ & $1.0\%$                \\
			\ac{MCP} efficiency & $0.997 \pm 0.003$ & $0.3\%$        \\
			Live time & $0.79806 \pm 0.00002$ & $0.002\%$                        \\
			
		\end{tabular}
	\end{ruledtabular}
	\label{table:Errors}
\end{table}

Analyzing the region of the contour plot surrounding the global minimum, we extract $68\%$ confidence intervals for the resonance energy and resonance strength of \meas{E_r = 679^{+2}_{-1}}{keV} and \meas{\wg = 120^{+50}_{-30}}{meV}. These quantities represent statistical uncertainties only. A number of sources of systematic uncertainty are also present, and are summarized in \tableref{table:Errors}. Note that the $0.17\%$ systematic uncertainty on the beam energy (c.f.~Section~\ref{sec:Exp}) is implicitly included since it is already folded into the quoted uncertainty on the stopping power. The resonance strength measurement is subject to systematic uncertainties related to each of the quantities in Table~\ref{table:Errors}, while the resonance energy measurement is affected only by the stopping power. Adding all of the relative uncertainties in quadrature, we arrive at the following resonance energy and strength values:
\begin{align*}
		E_r & = 679^{+2}_{-1}~\mathrm{(stat.)} \pm 1~\mathrm{(sys.)}~\mathrm{keV} \\
		\wg & = 120^{+50}_{-30}~\mathrm{(stat.)}^{+20}_{-60}~\mathrm{(sys.)}~\mathrm{meV}.
\end{align*}

The uncertainty due to potential background from reactions occurring on isobaric \nuc{38}{Ar} contamination in the beam was determined through a background measurement using a stable beam of pure \nuc{38}{Ar}, with a total ions on target of $(6.9 \pm 0.6) \ee{11}$. This measurement observed a single count near the edge of the expected recoil region, shown as the filled triangle in \figref{fig:e689coinc}(a). This count is likely a random leaky beam event based on its location in the \ac{IC} total energy vs.\ energy loss spectrum. This is demonstrated in \figref{fig:e689coinc}(f), which clearly shows that the suspected background event is well separated from the locus of \nuc{38}{K} recoils and is consistent with the locus of leaky beam events. Furthermore, the known properties of \nuc{38}{Ar $+p$} radiative capture imply that background from \nuc{38}{Ar} contamination is highly unlikely.  There are no known \nuc{38}{Ar $+p$} resonances within \meas{10}{keV} of the energies covered in the \ac{DRAGON} gas target \cite{Singh2006}. As a result, resonant capture is only possible through heretofore unknown proton-unbound states in the well-studied \nuc{39}{K} nucleus. Concerning direct capture, we calculate an estimated cross section of \meas{0.6}{nb} using the $S$-factor parametrization of \citeref{Longland2010}. Integrated across the length of the entire target, this results in an expected yield of only \meas{6 \ee{-5}}{recoils}.


Given the small likelihood that the single event observed in the measurement with pure \nuc{38}{Ar} beam is a genuine \rxn{38}{Ar}{p}{\gamma}{39}{K} recoil, we do not alter the \meas{\wg = 120}{meV}  central value extracted from our likelihood analysis. However, for a conservative estimate of the associated uncertainties, we recommend that the lower-bound systematic uncertainty include the possibility of unforeseen contamination arising from \rxn{38}{Ar}{p}{\gamma}{39}{K} reactions.  To calculate this uncertainty, we first determine an upper limit of $2.4$ events, or a yield of $3.4\ee{-12}$, in the pure \nuc{38}{Ar} beam measurement. We do this by applying the standard Rolke method \cite{Rolke2005493} to the single count observed in the recoil region. In the production runs with the \nuc{38}{K} radioactive beam (mixed with \nuc{38}{Ar} contamination), this translates into an upper limit of $13$ events. This upper limit is calculated assuming an  Ar/K ratio of $1.54$ in the production beam, determined by sending attenuated beam to the end of \ac{DRAGON} and fitting the individual Ar and K components in the \ac{IC} energy loss spectrum. Dividing by the $27$ observed recoil events, we arrive at a relative uncertainty of $50\%$. This uncertainty applies only to the lower limit on the resonance strength since the presence of background due to beam contamination can only reduce, never increase, the measured resonance strength. We emphasize that this procedure for determining a systematic uncertainty due to potential \nuc{38}{Ar} background is an \emph{ad hoc} adjustment, not one formulated from rigorous statistical methods. Overall, it provides a conservative estimate on the total systematic uncertainty applied to the resonance strength measurement.

The beam delivered to \ac{DRAGON} was also contaminated by isomeric \nuc{38m}{K} (\meas{E_x = 130}{keV}, $t_{1/2} = 924$~ms). The ratio of \nuc{38m}{K} to \nuc{38g}{K} was measured to be $7.1\ee{-2}$ at the ISAC yield station.  The yield measurements bypass the charge state booster, and hence some additional fraction of the isomers will decay before reaching \ac{DRAGON}. The delay between production and arrival at the \ac{DRAGON} target is dominated by the charge breeding time, which has been measured to be on the order of a few hundred milliseconds \cite{CSBTIME}. Taking a nominal delay time of \meas{400}{ms}, the \nuc{38m}{K}/\nuc{38g}{K} ratio would decrease to  $5.3\ee{-2}$ by the time the beam reaches the \ac{DRAGON} target.  Given the small fraction of \nuc{38m}{K} in the beam, no background from isomeric  capture is expected. 

\begin{figure*}
	\centering
	\includegraphics{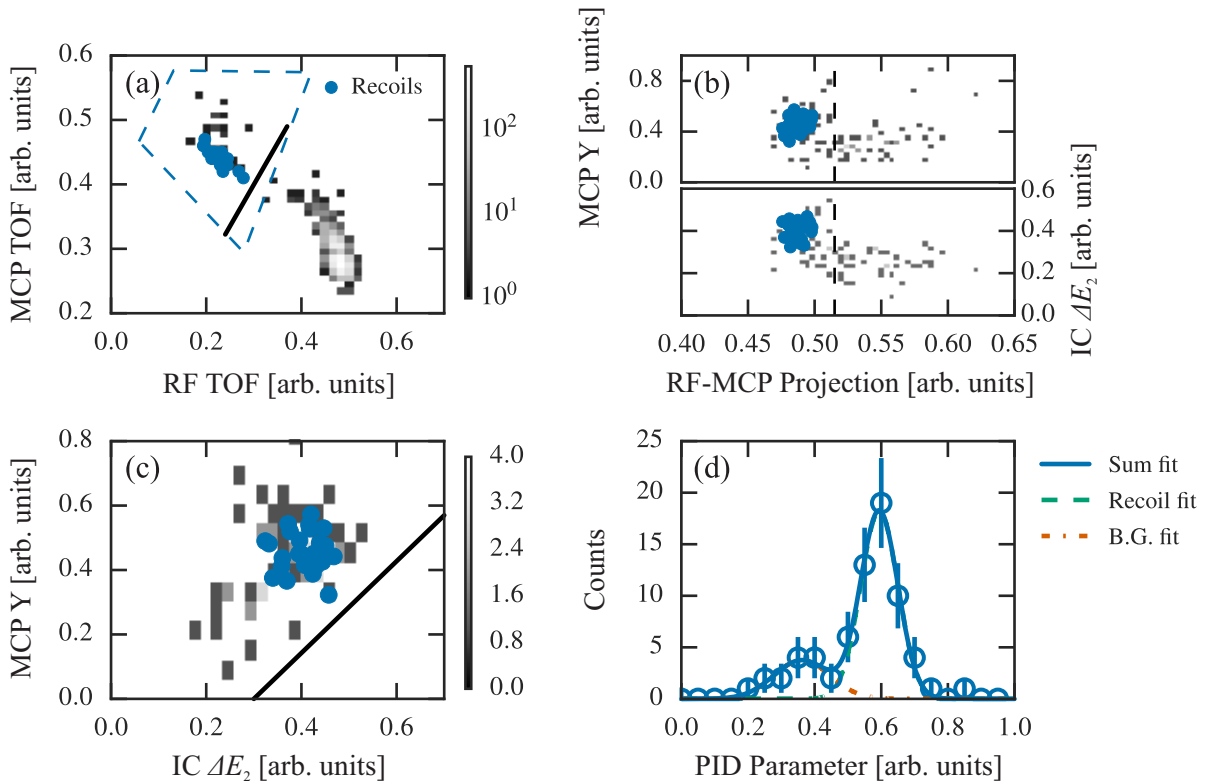} 
	\caption{
		Summary of the singles resonance strength analysis for the \ebeamC{} beam energy. In panels (a)~--~(c), the blue filled circles represent events already identified as recoils in the coincidence analysis, and the greyscale intensity maps represent all singles data. In panel (d), the open circles represent all  singles data, and the various curves represent fits as indicated in the legend. The solid black lines in panels (a) and (c) represent diagonal axes onto which the two-dimensional data are projected for subsequent analysis. The dashed black lines in panel (b) represent the cut placed on the ``RF-MCP Projection'' parameter. The full significance of each plot is explained in the main text.
}
	\label{fig:pidSingles}
\end{figure*}

\subsection{Singles Analysis}

In addition to the coincidence analysis of the \meas{689}{keV} resonance presented in Section~\ref{ss:Coinc}, we have also performed a separate extraction of the resonance strength using heavy-ion singles data alone. This analysis was \emph{guided by} the results of the prior coincidence analysis, i.e.\ regions of interest in various parameter spaces were identified by the location of coincidence recoils. However, the final quantitative cuts applied to the singles data were determined from the distributions of the singles parameters alone.
This singles analysis made use of the time difference between the incoming beam bunch (measured from the ISAC-I \ac{RFQ} signal) and the upstream \ac{MCP} to construct a separator TOF parameter without requiring prompt $\gamma$ rays. This analysis is summarized in the plots shown in \figref{fig:pidSingles}. Panel (a) shows the standard \ac{MCP} \ac{TOF} signal plotted vs.\ the RF--MCP \ac{TOF}, where the 27 events already identified as recoils in the coincidence analysis (represented by the blue filled circles) are tightly clustered in a narrow region of the plot. Continuing the analysis, we first set a gate on the entire upper-left region of the plot, which contains all of the coincidence recoils (the actual gate is included in the \figref{fig:pidSingles}(a) as the blue dashed line). We then project these events onto the solid black diagonal axis shown in the figure.

The new projected parameter (``RF-MCP projection'') is shown in the panel (b) of \figref{fig:pidSingles}, plotted vs.\ two separate parameters:
\begin{enumerate*}[label=\arabic*)]
	\item the $y$ position in the upstream \ac{MCP}, deduced from a resistive-anode readout scheme; and
	\item the energy loss in the third (most downstream) anode of the \ac{IC}.
\end{enumerate*}
In both cases, the confirmed recoil events are tightly clustered in a single region of the plot. To further separate recoil events from background, we place a one-dimensional cut on the ``RF-MCP projection'' parameter, including all events to the left of the black dotted line in the figure. For these events only, we then plot the \ac{IC} energy loss vs.\ the \ac{MCP} $y$ position, shown in \figref{fig:pidSingles}(c). Here, the singles events cluster into two distinct loci, with the confirmed recoil events falling entirely within the upper-right cluster. From this, we conclude that the singles events in the upper-right locus correspond to recoils, while the events in the lower-left locus correspond to background leaky beam events.

To quantify the overlap between the recoil and leaky beam regions in \figref{fig:pidSingles}(c), we project onto the diagonal axis indicated by the solid black line in the figure. This projection is shown in the \figref{fig:pidSingles}(d). The measured data (shown as open circles with error bars) are well-described by a double-Gaussian distribution (shown as dashed, dot-dashed, and solid lines, as indicated in the legend). The smaller Gaussian on the left of the figure corresponds to the estimated background distribution, and the larger Gaussian on the right of the figure corresponds to the recoil distribution. We take the true number of recoil events to be equal to the integral of the signal distribution, $52.0 \pm 8.2$. The uncertainty on this quantity comes from propagating the $1\sigma$ uncertainties on the individual fit parameters, which were calculated with MINUIT.



To calculate the singles resonance strength, we use the standard thick-target formula \cite{IliadisBook}, 
\begin{equation}\label{eq:WG}
\wg = {2 N_r \epsilon} / ({\eta N_b \lambda^2}),
\end{equation}
where $N_r = 52.0 \pm 8.2$ is the number of recoil events,
\meas{\epsilon = (3.95 \pm 0.14)\ee{-15}}{eV~cm$^2$} is the center-of-mass stopping power,
$\eta = 0.110 \pm 0.003$ is the heavy-ion  detection efficiency,
$N_b = (2.53 \pm 0.30)\ee{12}$ is the number of beam ions,
and \meas{\lambda = (3.513 \pm 0.005)\ee{-12}}{cm} is the center-of-mass deBroglie wavelength. Note that the heavy-ion detection efficiency includes the \ac{IC} efficiency of $0.913 \pm 0.003$. This  was not included in the heavy-ion efficiency used in the coincidence analysis since the \ac{IC} was not used to select coincidence events. The deBroglie wavelength  assumes a resonance energy of \meas{E_r = 679 \pm 2}{keV}, as extracted from our previous maximum likelihood analysis. The influence of this assumption is minor; calculating the resonance strength using the previous resonance energy of \meas{689\pm 10}{keV} increases the result by less than \meas{1}{meV}. The resulting resonance strength is \meas{\wg = 120 \pm 20}{meV} (statistical uncertainty only), which is in good agreement with our coincidence result of \meas{120^{+50}_{-30}}{meV} (the exact agreement of the central values should be considered fortuitous). The estimated singles systematic uncertainty is \meas{\pm 15}{meV}, calculated by propagating the uncertainties for the stopping power, number of beam particles, and detection efficiency.

In practice, the singles technique frequently results in a lower systematic uncertainty than the coincidence method since there is no need to estimate the $\gamma$-ray detection efficiency. This efficiency typically comes with a relative uncertainty of $10\%$ or greater, resulting from uncertainties in the \geant{3} simulation of the BGO array \cite{darioThesis}, as well as from unknown $\gamma$-ray decay schemes.
However, for reliable application of the singles technique, it is crucial that the full width of the resonance be contained within the gas target, to ensure that the thick-target approximation of the resonance strength formula is valid. In the future, technical advances will likely improve the ability to discern resonance positions based on only a handful of recoil events. With this capability, an off-center resonance would be spotted early on during a running period, and the beam energy could be adjusted accordingly. One example presently under development is the use of a fast-timing LaBr array for $\gamma$-ray detection. The fast timing properties of LaBr allow the resonance position to be deduced from the time difference between the detected $\gamma$ rays and the arrival of the corresponding beam bunch. Preliminary calculations and simulation work suggest that this method is more precise than the presently-employed $z$-position technique and may be applied to data sets with as few as ${\sim} 5$ confirmed recoils \cite{labrTiming}.

\section{Discussion}

\begin{figure}
	\centering
	\includegraphics{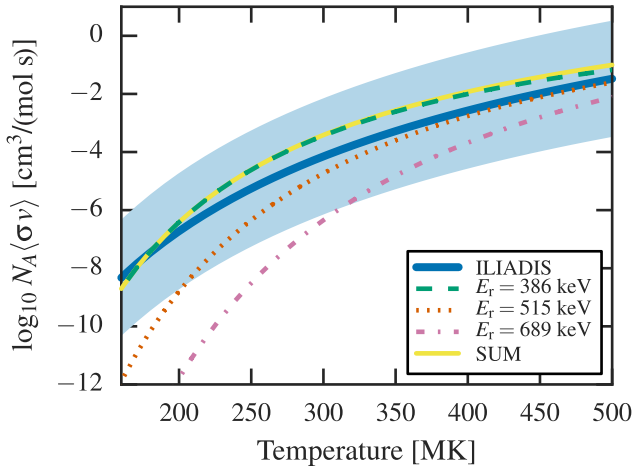}
	\caption{
		Updated \kpg{} reaction rate across the temperature regime covered by classical novae, calculated assuming the rate is dominated by the three $\ell = 0$ resonances studied in the present experiment. The contributions of the $386$ and $515$ keV resonances represent upper limits, while the ``689 keV'' resonance contribution represents our measured central value of \meas{120}{meV}. Also shown is the statistical model rate of \citeref{Iliadis2001} (``ILIADIS''), along with its associated uncertainties (shaded region).
}
\label{fig:Rates}
\end{figure}

\begin{table}
\centering
\caption{
	Calculated abundances for Ar, K, and Ca isotopes, from the NuGrid simulations explained in the text. Results from the Iliadis \emph{et al.}~sensitivity study, model ``S1'' (as well as model ``P2'', for \nuc{39}{K}) are included for comparison \cite{Iliadis2002}. The quantity $X_{\mathrm{rec}}$ represents the abundance calculated using the recommended \kpg{} rate from \citeref{Iliadis2001}, while $X_{100}$ and $X_{0.01}$ represent abundances calculated with the recommended rate multiplied by factors of $100$ and $0.01$, respectively. The quantities $X_{\mathrm{up}}$ and $X_{\mathrm{low}}$ represent abundances calculated with the experimental upper and lower limits presented in \citeref{PhysRevLett.116.132701}.
}
\begin{ruledtabular}
%
%
%
%
\begin{tabular}{ccccccc}
Nuclide	     & \Abr{100}{rec}	& \Abr{up}{rec} & \Abr{low}{rec} & \Abr{0.01}{rec} & \Abr{100}{0.01} & \Abr{up}{low} \\ \hline
\multicolumn{7}{c}{NuGrid} \\ \hline
\nuc{38}{Ar} & 0.066  & 0.57   & 1.4    & 1.4   & 1/21 &  1/2.5      \\
\nuc{39}{K}  & 3.4    & 2.1    & 0.14   & 0.094 & 36 &  15       \\
\nuc{40}{Ca} & 2.4    & 1.7    & 0.18   & 0.069 & 35 &  9.4      \\ \hline
\multicolumn{7}{c}{Iliadis \emph{et al.} \cite {Iliadis2002} } \\ \hline
\nuc{38}{Ar} (``S1'') & 0.057   & 0.60    & 1.4     & 1.4    &  1/25 &  1/2.3    \\ 
\nuc{39}{K} (``S1'')  & 3.4     & 2.0     & 0.19    & 0.059  &  58 & 11      \\
\nuc{39}{K} (``P2'')  & 9.5     & 2.6     & 0.17    & 0.070  &  136 & 15      \\
\nuc{40}{Ca} (``S1'') & 2.4     & 1.7     & 0.20    & 0.042  &  57 & 8.5     \\
\end{tabular}
\end{ruledtabular}
\label{table:NuGrid}
\end{table}

\begin{figure}
\centering
\includegraphics[width=3.4in]{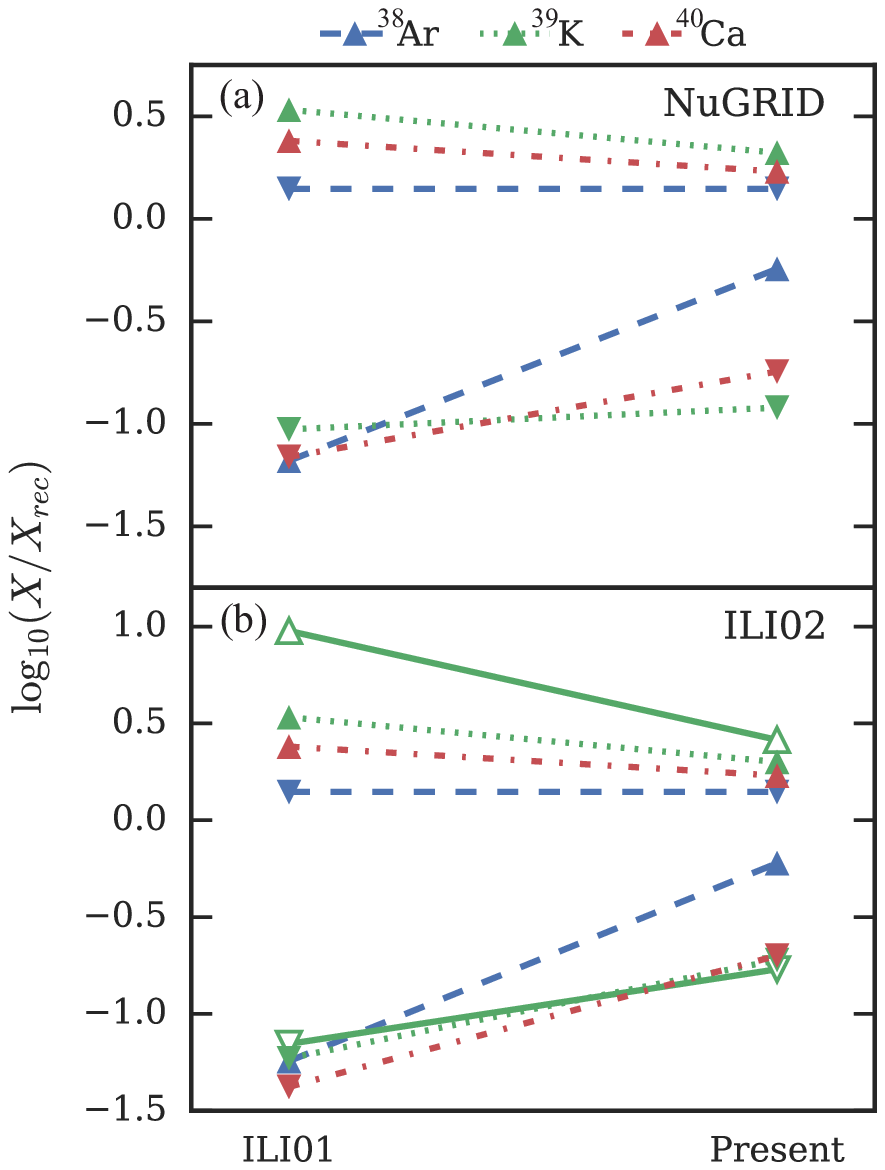}
\caption{
	Summary of the sensitivity study results presented in Table~\ref{table:NuGrid}. The various data points represent the logarithm of the ratio $X/X_{\textrm{rec}}$, where $X_{\textrm{rec}}$ is the predicted abundance of a given isotope taking the recommended \kpg{} rate from \citeref{Iliadis2001}, and $X$ is the predicted abundance of the same isotope taking the \kpg{} rate to be at the upper or lower limit of various uncertainty bands. For the points on the left of the figure (labeled ``ILI01''), $X$ is evaluated at the factor of 100~up~/0.01~down uncertainty limits given in \citeref{Iliadis2001}. For points on the right of the figure (labeled ``Present''), $X$ is taken from the uncertainty band established in \citeref{PhysRevLett.116.132701}. In all cases, up-turned triangles represent $X$ calculated at the upper limit of the uncertainty band and down-turned triangles represent $X$ at the lower limit. The various dashed, dotted, and dot-dashed lines represent abundance calculations for \nuc{38}{Ar}, \nuc{39}{K}, and \nuc{40}{Ca} as indicated in the legend at the top of the figure. Panel~(a)  shows the results of the present NuGRID sensitivity study, while panel~(b)   shows the results of the Iliadis~\textit{et al}.\ sensitivity study \cite{Iliadis2002}. In panel~(b), results from both the ``S1'' and ``P2'' nova models are displayed for \nuc{39}{K}. The filled green triangles connected by the dotted line represent results of the ``S1'' model, while the open triangles connected by solid lines represent the results of the ``P2'' model.
}
\label{fig:abundances}
\end{figure}

As discussed in \citeref{PhysRevLett.116.132701}, the present measurements place significant constraints on the overall \kpg{} reaction rate at nova temperatures. This is demonstrated in \figref{fig:Rates}, which shows the calculated rate~vs.~temperature curves for the three presently reported resonances, along with their sum. Assuming the astrophysical rate is dominated by these three resonances, the lower curve (the nominal \meas{689}{keV} resonance) sets a lower limit on the astrophysical rate, while the sum sets an upper limit. For comparison, the recommended rate from Iliadis \textit{et al.}, along with the factor $100$ up/down uncertainty band, is also included in the figure. 
At peak temperatures for ONe nova burning, $T \simeq 0.4~${}GK, the total uncertainty has been reduced from a factor of $10^4$ to a factor of ${\sim} 40$. Applying these new, experimentally based, limits to the model predictions of the Iliadis \emph{et al}.\ sensitivity study  \cite{Iliadis2002} results in a reduction of overall uncertainties on \nuc{38}{Ar}, \nuc{39}{K}, and \nuc{40}{Ca} production in ONe novae from respective factors of ${\sim} 25,$ $136,$ and $57$ to factors of ${\sim} 2,$ $18,$ and $9$. Note for these calculations, the nucleosynthesis models which maximized the sensitivity to the \kpg{} rate were used. For \nuc{38}{Ar} and \nuc{40}{Ca} this corresponds to the ``S1'' model (\meas{T_{\mathrm{peak}} = 418}{MK}), while for \nuc{39}{K} the ``P2'' model (\meas{T_{\mathrm{peak}} = 356}{MK}) was used.

In order to investigate the dependence of these sensitivity results to specific nova models, we have performed a separate \kpg{} sensitivity study based on an independent calculation performed with the NuGrid package, using the single-zone ``post processing network'' (ppn) code  \cite{Denissenkov11082014}. The initial conditions of the calculation are a $1.3$ solar mass white dwarf with a temperature of \meas{7.0}{MK}. The white dwarf composition is given by the ``Denisenkov'' model, evolved using the Modules for Experiments in Stellar Astrophysics (MESA) code \cite{2011ApJS..192....3P}. The accretion rate is $10^{-11}$ $M_{\odot}/\mathrm{yr}$, and the composition of the accreted material is assumed to be solar. The peak temperature of the model outburst is \meas{408}{MK}, similar to the \meas{418}{MK} peak temperature of the S1 model from \citeref{Iliadis2002}. A complete description of the parameters going into the NuGrid calculation can be found in \citeref{Denissenkov11082014}. The results of the NuGrid sensitivity study are summarized in \tableref{table:NuGrid}, with results from the Iliadis \emph{et al.}~study included for comparison. The recommended rate utilized for this analysis is identical to the one presented in \citeref{Iliadis2001}. Overall, the predictions of the NuGrid and the S1 models are rather consistent, with agreement to within a factor of two in all cases.

Our measurements and sensitivity analyses indicate that the \kpg{} rate is not a likely source of significant over- or under-production of \nuc{38}{Ar}, \nuc{39}{K} or \nuc{40}{Ca} in novae (relative to solar abundances). Hence the over-production of Ar and Ca observed in the spectra of nova ejecta \cite{Pottasch1959, Andrea1994, Arkhipova2000, Evans2003} remains unexplained. We encourage more extensive sensitivity studies and multi-zone model calculations to investigate the source of this anomaly.

It should be noted that the present  results intrinsically depend on the veracity of previous transfer reaction studies, which have established the \nuc{39}{Ca} level scheme in the $E_x = 6$--$7$~MeV region. If the spins or level energies established from these studies are incorrect or incomplete, the present experiment may have neglected to cover the most important resonance energy windows for astrophysics. For this reason, we encourage future high-resolution transfer reaction studies that are targeted specifically at measuring the properties of potential astrophysical proton capture resonances in \nuc{39}{Ca}.

Although the present measurements are not directly sensitive to the spins of the measured resonances, we can still infer properties of the resonances in question based on the measured strengths. For this, we use the standard formula for the resonance strength \cite{IliadisBook},
\begin{equation}
\label{eq:ResStrengthFormula}
	\wg = \frac{2 J_r + 1}{\left(2 J_p + 1 \right) \left(2 J_{{}^{38}\mathrm{K}} + 1 \right)}
\frac{\Gamma_\gamma \Gamma_p}{\Gamma_\gamma + \Gamma_p}, 
\end{equation}
where $J_r = \nicefrac{5}{2}$, $J_p = \nicefrac{1}{2}$, and $J_{{}^{38}\mathrm{K}} = 3$ are the respective spins of the resonance, proton, and \nuc{38}{K}; and $\Gamma_\gamma$ and $\Gamma_p$ are the respective $\gamma$-ray and proton partial widths of the resonance.  Assuming a ``hard'' upper limit on the proton spectroscopic factor for unbound states of $C^2 S \leq 0.1$ and the measured strength value of \meas{\wg = 120}{meV} for the \meas{679\pm 2}{keV} resonance, we calculate an upper limit on the mean $\gamma$-decay lifetime for this state of \meas{\tau \leq 2.2}{fs}. For shorter lifetimes, as $\tau \rightarrow 0$, $\Gamma_\gamma / (\Gamma_\gamma + \Gamma_p) \rightarrow 1$,
and we calculate a lower limit on the spectroscopic factor of $C^2 S \geq 0.0055$. For the \meas{515 \pm 10}{keV} resonance, the $90\%$ confidence level (CL) upper limit on the strength of \meas{18.4}{meV} sets an upper limit on the lifetime of \meas{\tau \leq 12}{fs} (again taking the ``hard'' upper limit on the spectroscopic factor at $C^2 S = 0.1$). 
For short lifetimes, $\Gamma_\gamma / (\Gamma_\gamma + \Gamma_p) \simeq 1$, we calculate an upper limit on the spectroscopic factor of $C^2 S \leq 0.022$. For the \meas{386 \pm 10}{keV} resonance, the calculated upper limit on the lifetime is \meas{\tau \leq 38}{fs} (again taking the measured $90\%$ upper limit on the strength of \meas{\wg \leq 2.54}{meV} and the ``hard'' spectroscopic factor limit of $0.1$).  For short lifetimes satisfying $\Gamma_\gamma / (\Gamma_\gamma + \Gamma_p) \simeq 1$, we calculate an upper limit on the spectroscopic factor of $C^2 S \leq 0.066$. We emphasize that these limits are simply ``back of the envelope'' calculations and not intended to set any rigid limits on the single-particle properties of the resonances in question. 

To summarize, we have performed the first ever direct measurement of the \kpg{} reaction, focusing on the three potential $\ell = 0$ resonances within the Gamow Window for classical novae, whose energies have been determined previously to be \meas{386 \pm 10}{keV}, \meas{515 \pm 10}{keV}, and \meas{689 \pm 10}{keV}. For the highest-energy resonance, we observed a clear \nuc{39}{Ca}--$\gamma$ coincidence signal consisting of $27$ events. We performed a two-dimensional likelihood analysis on the position distribution of the measured $\gamma$-rays to extract a resonance strength and energy of $\wg = 120^{+50}_{-30}\mathrm{(stat.)}^{+20}_{-60}\mathrm{(sys.)}~\mathrm{meV}$ and $E_r = 679^{+2}_{-1}\mathrm{(stat.)} \pm 1\mathrm{(sys.)}~\mathrm{keV}$, respectively. The quoted systematic uncertainties are conservative and include the possibility of background events arising from stable \nuc{38}{Ar} beam contamination. We also performed a separate analysis of \nuc{39}{Ca} singles data and extracted a resonance strength of $\wg = 120 \pm 20\mathrm{(stat.)} \pm 15\mathrm{(sys)}~\mathrm{meV}$, consistent with the coincidence result. For the lower two resonances, we observed no events consistent with recoils and used a profile likelihood technique to extract $90\%$ CL upper limits on the resonance strengths of \meas{2.54}{meV} and \meas{18.4}{meV} for the lower and middle resonances, respectively. Based on these measurements we have established new recommended upper and lower limits for the \kpg{} reaction rate which reduce uncertainties at peak nova temperatures ($T_{9} \sim 0.4$) from a factor of $10^4$ to a factor ${\sim} 40$. Incorporating these new limits into two separate nova model calculations we find that the uncertainties on the predicted abundances of \nuc{38}{Ar}, \nuc{39}{K}, and \nuc{40}{Ca} are reduced to a factor of $15$ or below in all cases.


%

%

\acknowledgements{}
	The authors are grateful to the ISAC operations team and the technical staff at TRIUMF for their support during the experiment, in particular F.~Ames for dedicated operation of the \ac{ECRIS} charge state booster. We also thank R.~Wilkinson for calculations of proton single-particle widths for the three resonances measured in this work.	TRIUMF's core operations are supported via a contribution from the federal government through the National Research Council of Canada, and the Government of British Columbia provides building capital funds. DRAGON is supported by funds from the National Sciences and Engineering Research Council of Canada. Authors from the Colorado School of Mines acknowledge support from the Department of Energy, grant DE-FG02-93ER-40789. The U.K. authors acknowledge support by STFC.
    
The authors acknowledge P.~Denisenkov for assistance in running the NuGrid code and for fruitful discussions. Support for NuGrid is provided by the National Science Foundation through grants PHY~02-16783/PHY~08-22648 and PHY-1430152, which fund the Joint Institute for Nuclear Astrophysics (JINA) and the JINA Center for the Evolution of the Elements, respectively. NuGrid support is also provided by the European Union through grant MIRG-CT-2006-046520. The NuGrid collaboration uses services of the Canadian Advanced Network for Astronomy Research (CANFAR) which in turn is supported by CANARIE, Compute Canada, University of Victoria, the National Research Council of Canada, and the Canadian Space Agency.



%

\end{document}